# The double tetrahedron structure of the nucleus

Jozsef Garai
Department of Earth Sciences
Florida International University
University Park, PC-344
Miami, FL 33199
Ph: 305-348-3445
Fax: 305-348-3070
E-mail: jozsef.garai@fiu.edu

**Expanding a double tetrahedron formation of equal spheres arranged in fcc structure correlation between the positions of the nucleons and quantum numbers has been detected. The number of protons in the structure is not simply consistent with all the corresponding quantum numbers but also bears the same physical meaning as in quantum mechanics. The detected correlations between lattice positions of the protons and quantum numbers raise the possibility the solid nuclear structure might be able to provide an explanation for the single particle properties of the nuclei.**

The nuclei has been identified by the scattering experiments of Hans Geiger and Ernest Marsden, carried out under the supervision of Rutherford[1]. Ever since then this tiny center of the atom has been targeted by better and better penetrating probes to reveal its physics. Despite the great amount of knowledge has been gathered about the properties of the atomic nuclei even the phase of the nucleons remains an enigma.

The developed models, shell[2-6], liquid drop[7-8] and cluster[9] are assume a gas, liquid and semi-solid phase for the nuclei respectively. The observed saturation properties of the nuclear forces, the low compressibility of the nucleus and the well defined nuclear surface are consistent with a liquid phase, the independent quantum characteristics of the nucleons can be explained by assuming a gas phase for the nuclei while the clustering of alpha particles is best explained with a semi-solid phase. Each of these models is able to describe very successfully certain selected properties of the nuclei; however, the phases are mutually exclusive, therefore, neither of the model has any chance to provide a comprehensive description. The solution has to come from a new direction and the only remaining possibility left is the solid phase, which will be considering in details here.

Solid phase has not been considered as a viable option for many decades because of the uncertainty principle and the lack of diffraction. In the 1960s the discovery of quarks and neutron star researches satisfactorily answered these objections and opened the door for solid nuclear structures. The proposed models assume that the protons and the neutrons have same size; the individual nucleon has a spherical shape; protons and neutrons are alternating; and the nucleons are arranged in a closest packing crystal structure. All of these assumptions are reasonable. The radii of protons and the neutrons differ only slightly[10]. The same proton and neutron magic numbers indicates same structural development for both protons and neutrons. The same arrangement should result in an alternating proton and neutron array. The spheres most likely will be utilizing the available space in the most efficient way resulting in a closest packing arrangement. The most promising agreement with different nuclear properties have been find for face-centered cubic arrangement[11-17], which structure is favored by most of the researchers.

The solid phase description of the nuclei able to explain the collective and clustering properties of the nuclei, however, the explanation of individual properties has been remained a challenge for the model. The only solid phase property which might be able to explain the individual characteristics of the nuclei is the lattice positions of the nucleons. Significant effort has been made to find correlation between lattice positions and quantum numbers with limited success for fcc structure[18-19]. No correspondence between lattice positions and quantum states have been find for hexagonal closest packing, body centered cubic, and simple cubic packing[20]. Every solid model, probably rooting from the shell model, expanded the investigated the structure spherically. However the structure of fcc consist of tetrahedron and octahedron units; therefore, symmetry investigations should primarily concentrate on the extension of these arrangements.

The first stage in the nuclear structural development is marked by the end of the first period of the periodic system. Assuming closed packing arrangement for the nucleons, the two protons and two neutrons in the structure of the ${}^{4}_{2}He_2$ nuclei should form a tetrahedron; therefore, the tetrahedron expansion in fcc structure will be considered in details. Calculations of potential model, constrained by the hadron spectrum for the confinement of the relativistic quark[21-22] and colored quark exchange model[23] are also consistent with a tetrahedron formation of the ${}^{4}_{2}He_2$ nuclei.

Tetrahedron formation of equal spheres arranged in fcc packing can be expanded by adding layers of equilateral triangles packed in two dimensional closest packing. Starting with one sphere and increasing the length of the side of a triangle by one sphere the number of nucleons in each triangle will be 1, 3, 6, 10, 15, 21, 28, and 36 (Fig. 1/A) Placing one layer on to the other the number of spheres in two consecutive layers will be 4, 16, 36, and 54 giving the number of protons in these double layers to 2, 8, 18, and 32. These numbers are identical with the number of possible states of the principle quantum numbers. If the layers are added to the outer sides of the tetrahedron then a shell like structures can be formed (Fig. 1/D).

If one assumes that the outer layers of the double tetrahedron correspond to the radius [R(r)] then correlations between the number of protons and the other two quantum numbers might also exist. The angular momentum quantum number (l) corresponds to $[\Theta(\theta)]$, while the magnetic quantum number $(m_l)$ correspond to $[\Phi(\phi)]$.

Investigating the number of spheres in the vertexes of the tetrahedron it has been found that if the new shell covers two sides of the tetrahedron then the number of protons in one layer of the outer planes is the same as the number of states determined by the angular momentum quantum numbers. The number of different proton positions in one layer is the same as the number of magnetic quantum numbers (Fig. 2). In order to cover two sides of every vertexes of the tetrahedron a double tetrahedron has to be formed around a core tetrahedron. The core tetrahedron consist four nucleons. The arrangement of nucleons in the fully developed double tetrahedrons has been shown in Fig. S 1. The ideal coordinates of these noble gas nucleons are given in Tab. S 1.

The core tetrahedron consist four nucleons. This double tetrahedron has 2-8-8-18-18-32-32 proton cycles which is consistent with the periodicity of the periodic system. The periodic system of the elements exhibits a K-L-L-M-M-N-N development of the shells (Fig. S 2). The double periodicity in the L, M, and N shells is supported by all the characteristic features of the elements (Fig. S 3). Quantum mechanics explains the double periodicity by the completion of shells and subshells; however, there is no explanation why certain shell and subshell numbers are outstanding among the many numbers corresponding to the completion of the shells and subshells.

Counting the total number of protons in the same layers of the vertexes of the double tetrahedron it can be shown that the proton numbers are consistent with the multiplicities (Fig. 3).

If symmetrical positions are available in the structure then protons and neutrons predictable will occupy those positions primarily. Employing this assumption the most likely charge distribution of few nuclei in the first three periods have been predicted (Fig. S 1). The three dimensional images of these charge distributions along with the structure of the noble gases has been plotted[24] on Fig. 4.

Other nuclear properties supporting a double tetrahedron nuclear structure are the followings. The four particle symmetry cycle, which is the characteristic of tetrahedron formation, is consistent with the zero ground state angular momentum observed for even-even number nucleuses. The observed fissions can be explained by breaking the tetrahedron formation into two parts along a plane of the fcc structure. The elongated non-spherical shape of the nuclei[25-26], observed mainly in the middle of the periods, is consistent with the development of a double tetrahedron nuclei. The so called skin thickness part of the nuclei is consistent with a density or charge distribution of nucleons arranged in the four vertexes of the tetrahedron[27]. The detected laminar clustering of the nucleons[28] is consistent with a crystal structure.

Expanding a four nucleons core tetrahedron to a double tetrahedron formation it has been found that the number of protons in the outer two layers of the double tetrahedron corresponds to the principle quantum numbers, the number of protons in a layer of these shells corresponds to the angular momentum quantum number, while the number of different proton positions in one layer corresponds to the magnetic quantum number. These correlations are consistent with the physical meaning of the quantum description; therefore, a random coincidence should be excluded. The detected correlations raised the possibility that the double tetrahedron crystalline model might provide the first comprehensive description of the nuclei by combining the individual particle, collective, and clustering characteristics of the nuclei.

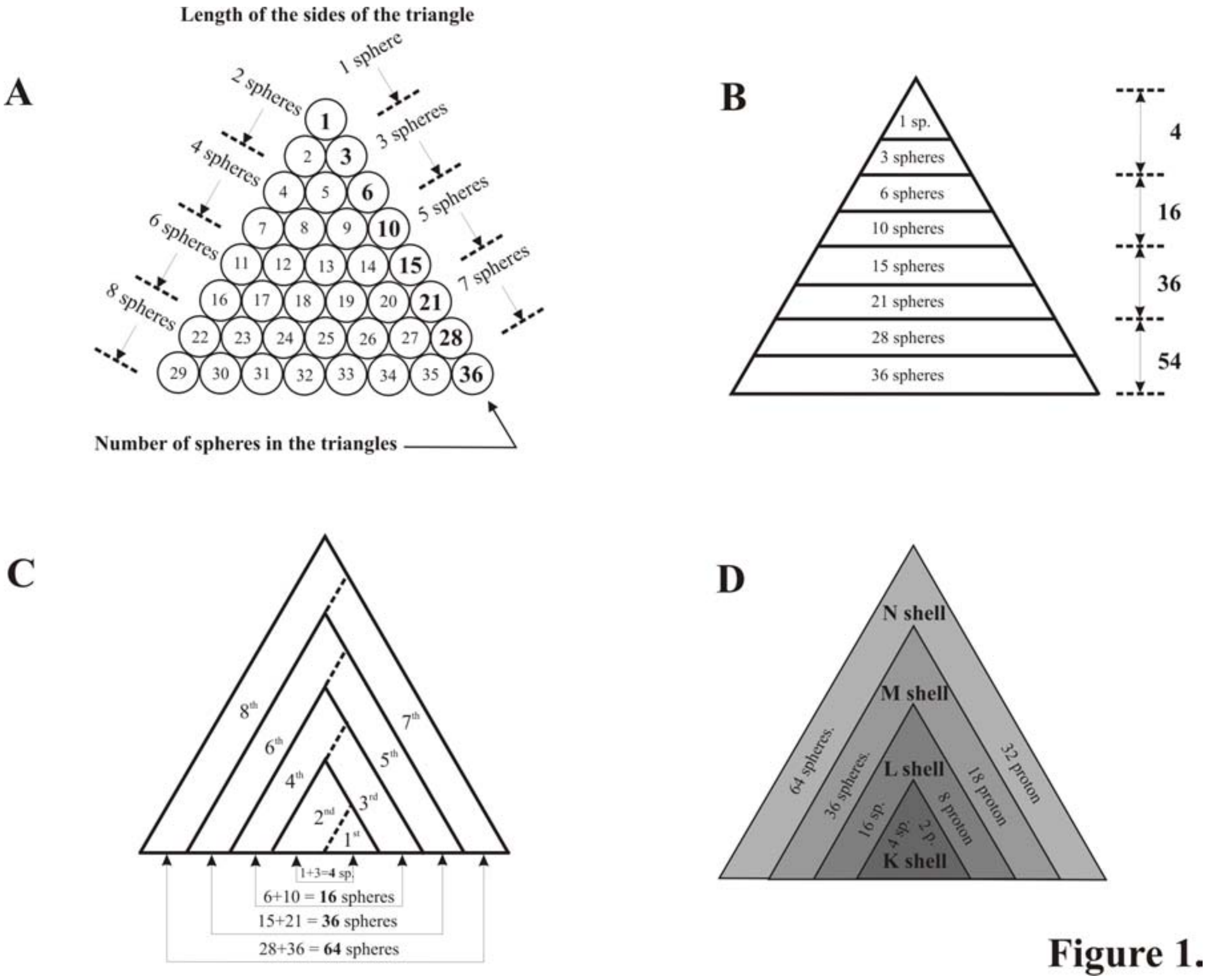

Figure 1.

**Fig. 1** Representing both protons and neutrons with equal spheres and arranging them in fcc structure the number of protons in the outer layers of a tetrahedron formation is the same as the number of possible states of the principle quantum numbers.

**(A)** The number of spheres in a two dimensional closed packing arrangement forming equilateral triangles.

**(B)** The number of spheres in two consecutive layers of the tetrahedron formation

**(C)** The same tetrahedron formation can be developed by adding the new layers to alternate sides.

**(D)** Assuming a proton-neutron ratio of one the outer layers of the tetrahedron contain the same number of protons as predicted by quantum theory.

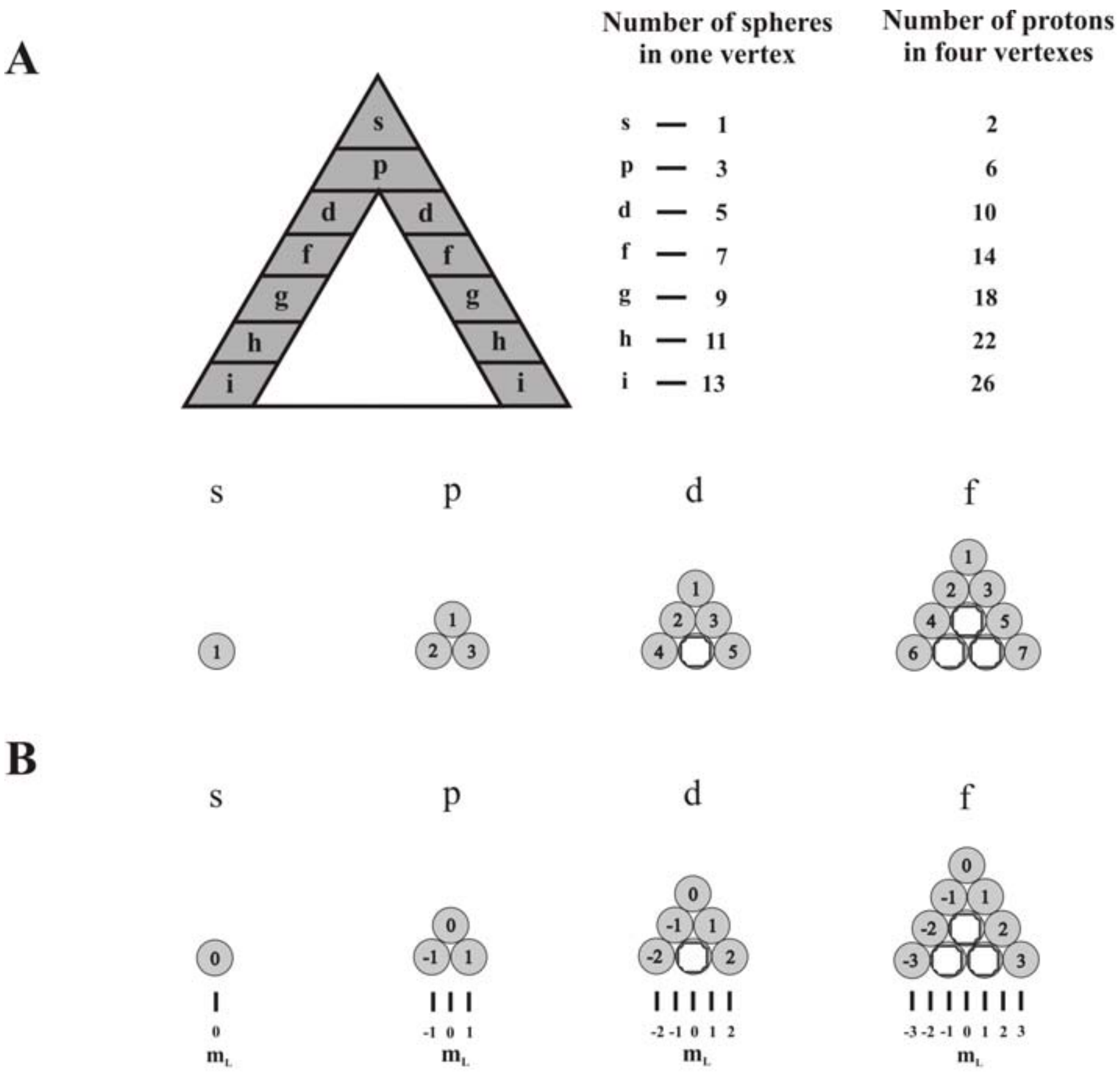

**Figure 2.**

**Fig. 2** If a double tetrahedron has been formed around a core tetrahedron, containing four spheres, then the number of protons in one layer of the outer shell is equivalent with the number of states of the corresponding subshells. The number of different positions of protons in one layer of the shell is the same as the number of magnetic quantum numbers. The proton number has been calculated by multiplying the number of spheres in one vertex of the tetrahedron by two. There are four vertexes, and every other sphere represents a proton.

**(A)** Number of spheres in one layer of a vertex of the double tetrahedron

**(B)** Number of the different proton positions in one layer of a vertex of a double tetrahedron.

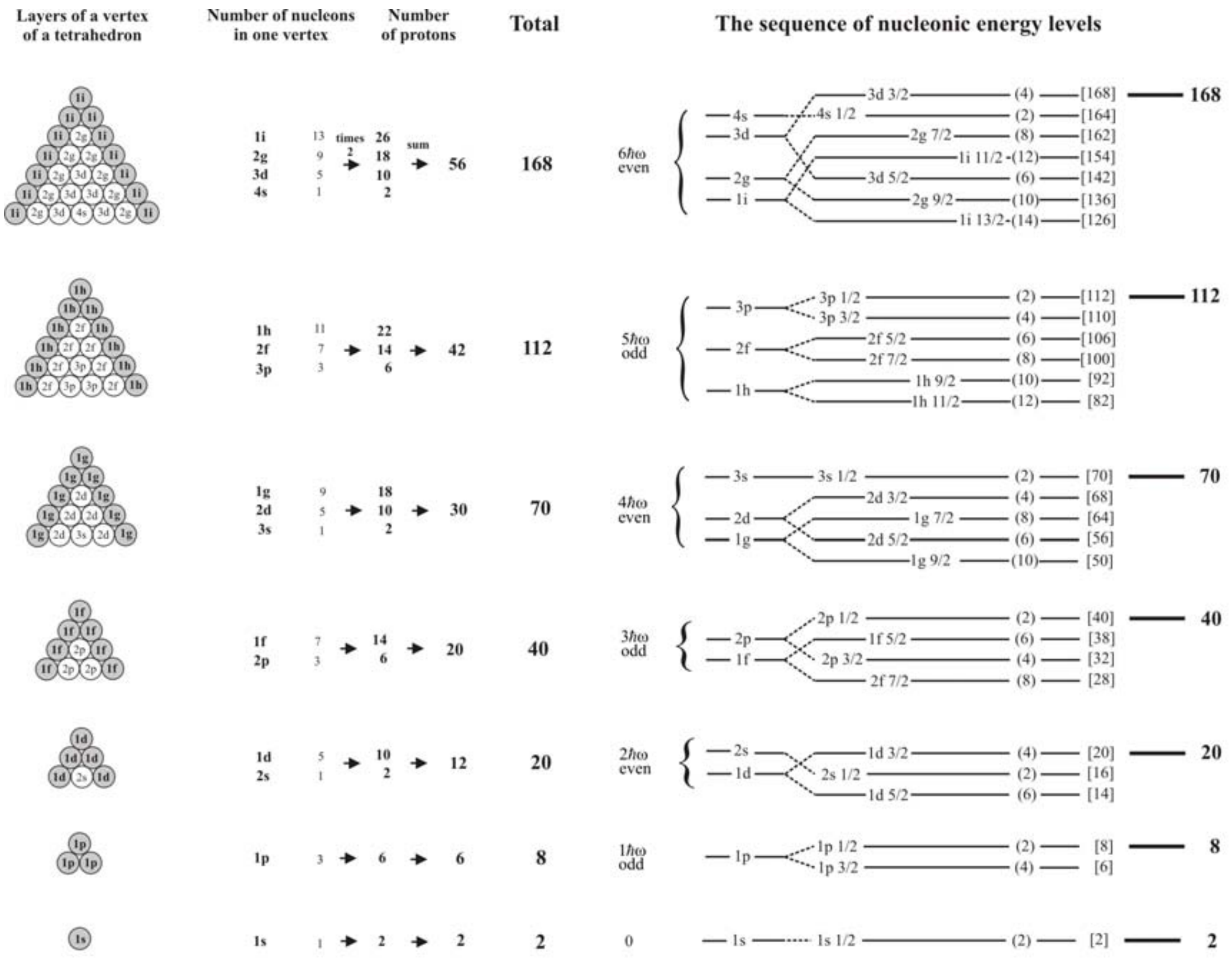

**Figure 3.**

**Fig. 3** The total number of protons in the layers of the vertexes of the tetrahedron is consistent with multiplicities. The number of spheres in one layer of a vertex of the tetrahedron is shown. The number of protons has been determined as it is given in Fig. 2. For comparison the corresponding energy states are also shown.

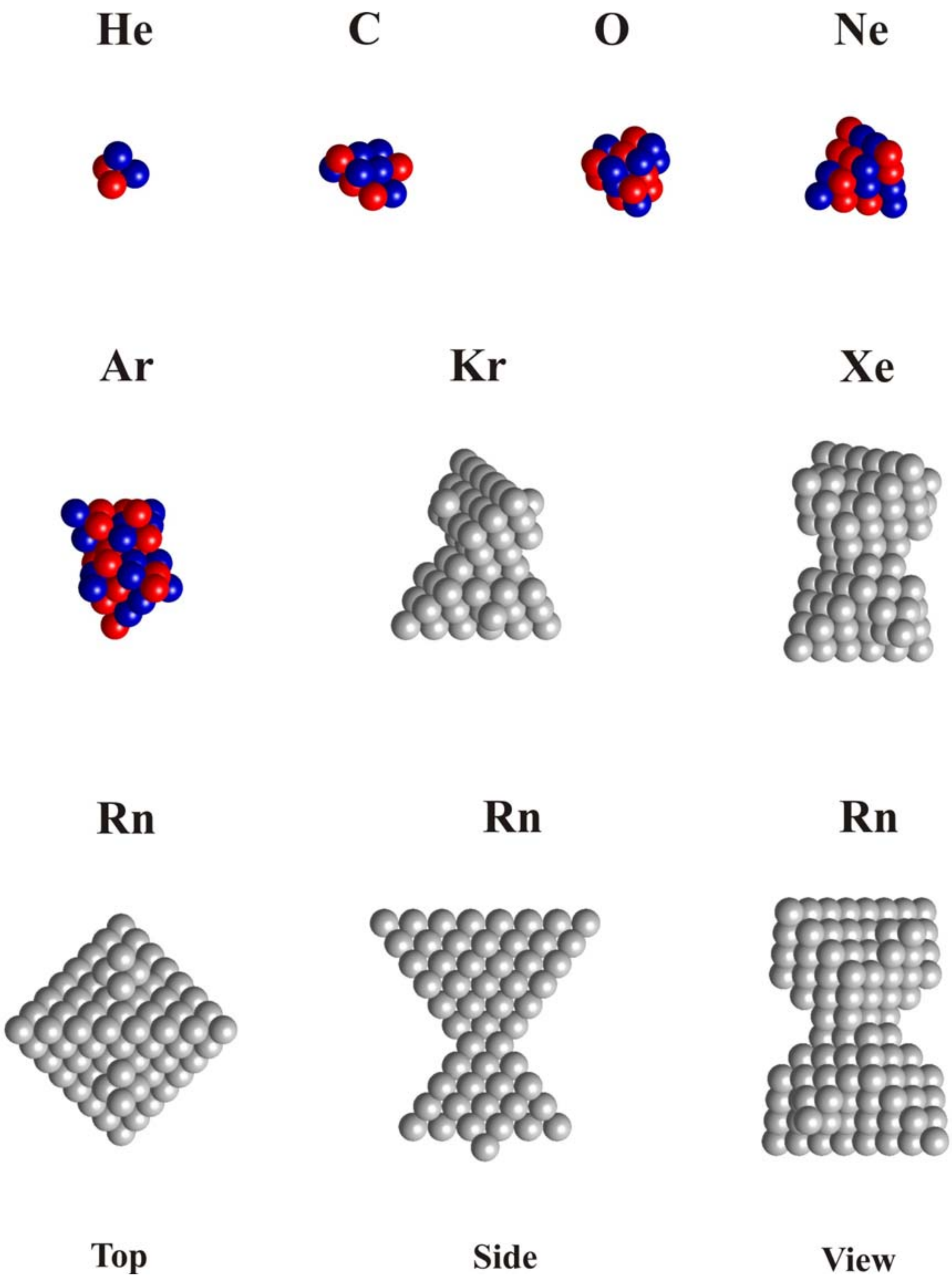

**Fig. 4** The 3D images of the structures of the even proton neutron number nucleons in the second period and the fully developed nuclear structures of the noble gases has been plotted. The red and blue spheres represent protons and neutrons respectively. In structures, where the charges are not predicted, grey sphere has been used to represent a nucleon.

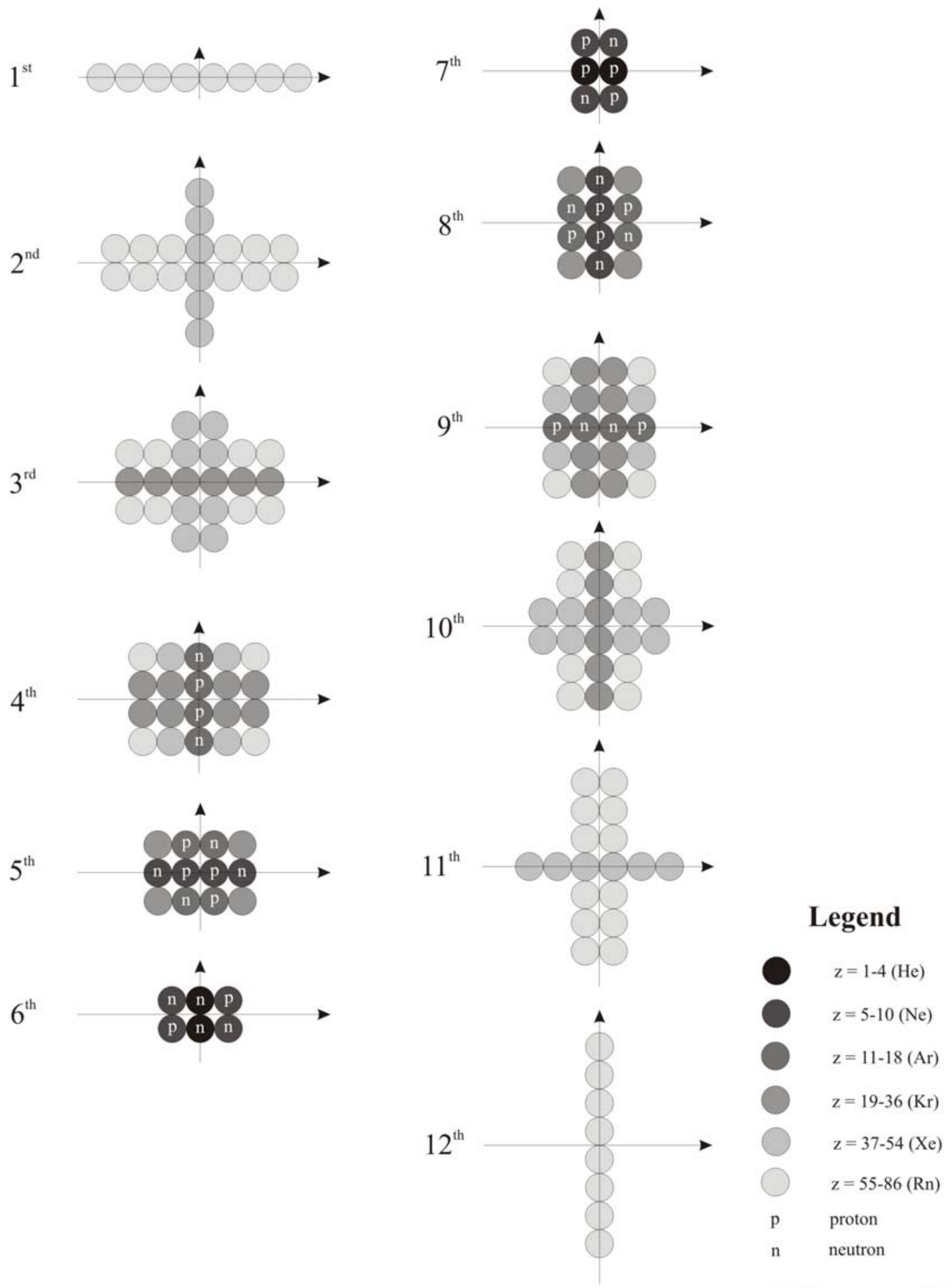

Figure S1

**Fig. S 1** The arrangement of the nucleons in the fully developed double tetrahedron structures. The numbered layers are displayed from each other by the distance of $d\sqrt{\frac{2}{3}}$, where d is the diameter of the nucleons.

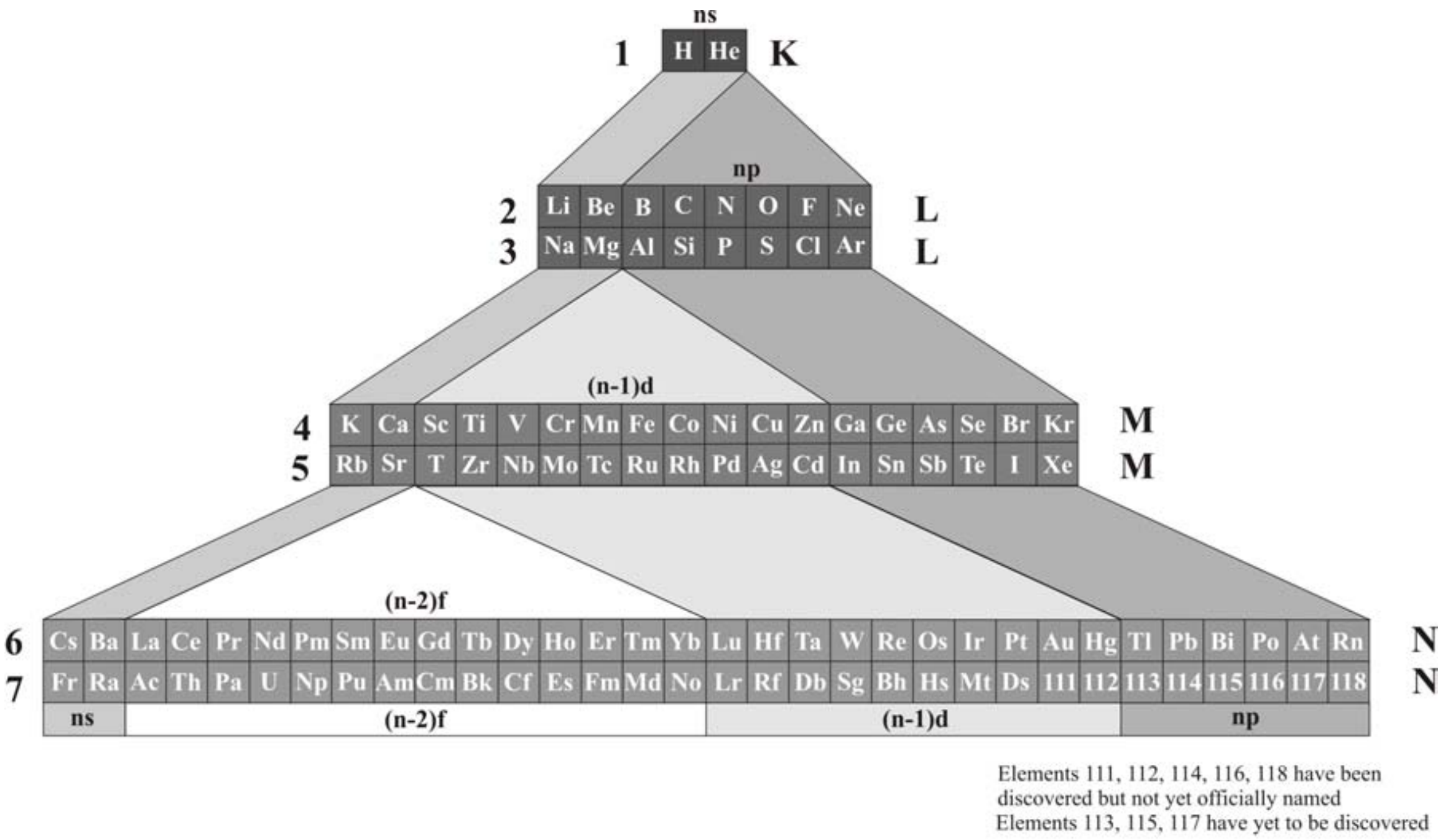

**Figure S2**

**Fig. S 2** The periods in the periodic system, except for the first period, are repeated. The length of the periods is consistent with quantum predictions; however, development of the shells is not K-L-M-N-O-P-Q but rather K-L-L-M-M-N-N.

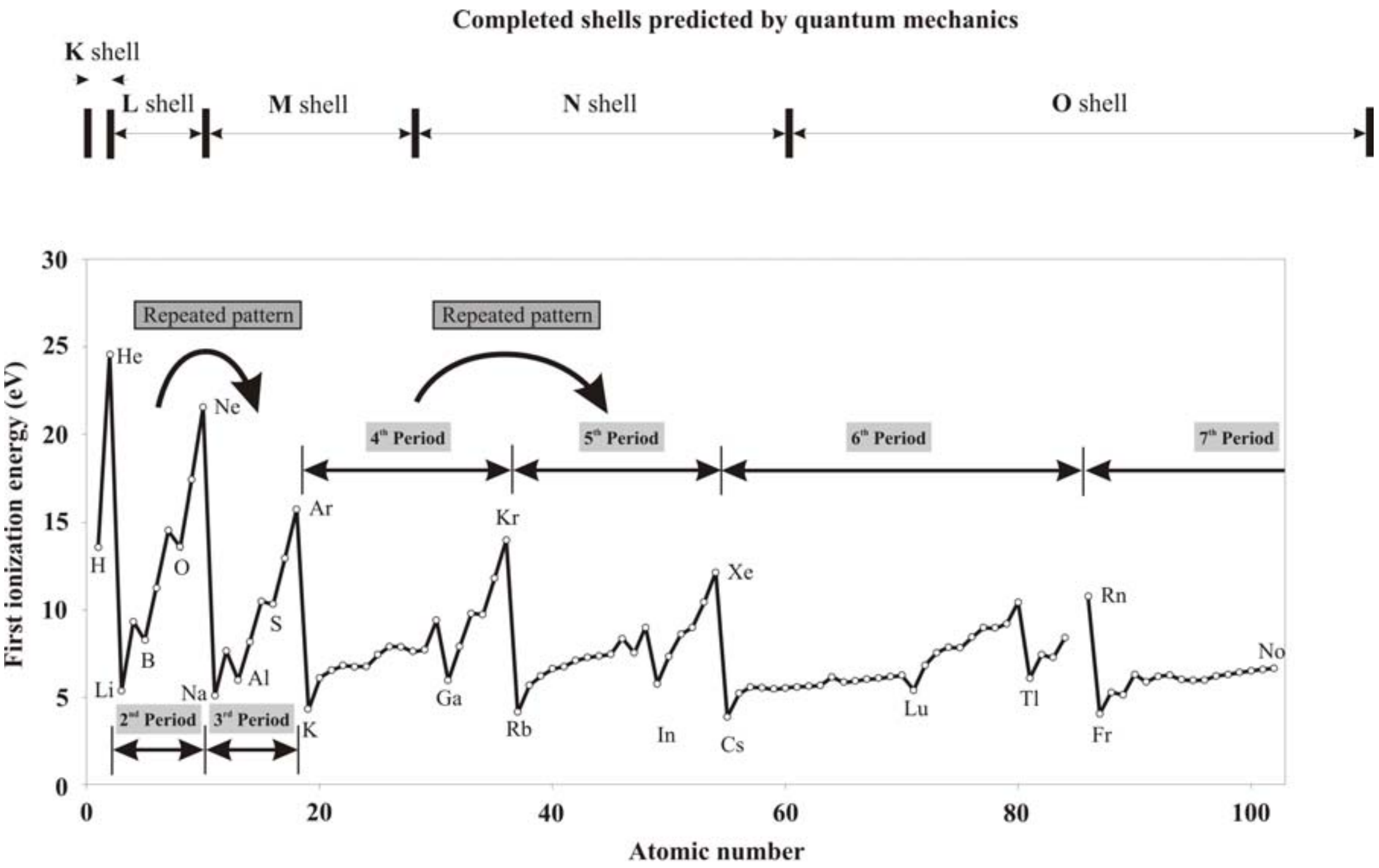

**Figure S3**

**Fig. S 3** The double periodicity of the L, M, and N shells is visibly present in any characteristic features of the elements. As an example the first ionization energy of the element has been plotted.

| Nucleon number | Corresponding Period | Sub-shell | X | Coordinates Y | Z |
|---|---|---|---|---|---|
| 1 | 6 | s | -3.5 | 0.0 | -3.889 |
| 2 | 6 | p | -2.5 | 0.0 | -3.889 |
| 3 | 6 | d | -1.5 | 0.0 | -3.889 |
| 4 | 6 | f | -0.5 | 0.0 | -3.889 |
| 5 | 6 | f | 0.5 | 0.0 | -3.889 |
| 6 | 6 | d | 1.5 | 0.0 | -3.889 |
| 7 | 6 | p | 2.5 | 0.0 | -3.889 |
| 8 | 6 | s | 3.5 | 0.0 | -3.889 |
| 9 | 6 | p | -3.0 | -0.5 | -3.182 |
| 10 | 6 | p | -3.0 | 0.5 | -3.182 |
| 11 | 6 | d | -2.0 | -0.5 | -3.182 |
| 12 | 6 | d | -2.0 | 0.5 | -3.182 |
| 13 | 6 | f | -1.0 | -0.5 | -3.182 |
| 14 | 6 | f | -1.0 | 0.5 | -3.182 |
| 15 | 5 | s | 0.0 | -2.5 | -3.182 |
| 16 | 5 | p | 0.0 | -1.5 | -3.182 |
| 17 | 5 | d | 0.0 | -0.5 | -3.182 |
| 18 | 5 | d | 0.0 | 0.5 | -3.182 |
| 19 | 5 | p | 0.0 | 1.5 | -3.182 |
| 20 | 5 | s | 0.0 | 2.5 | -3.182 |
| 21 | 6 | f | 1.0 | -0.5 | -3.182 |
| 22 | 6 | f | 1.0 | 0.5 | -3.182 |
| 23 | 6 | d | 2.0 | -0.5 | -3.182 |
| 24 | 6 | d | 2.0 | 0.5 | -3.182 |
| 25 | 6 | p | 3.0 | -0.5 | -3.182 |
| 26 | 6 | p | 3.0 | 0.5 | -3.182 |
| 27 | 6 | d | -2.5 | -1.0 | -2.475 |
| 28 | 4 | s | -2.5 | 0.0 | -2.475 |
| 29 | 6 | d | -2.5 | 1.0 | -2.475 |
| 30 | 6 | f | -1.5 | -1.0 | -2.475 |
| 31 | 4 | p | -1.5 | 0.0 | -2.475 |
| 32 | 6 | f | -1.5 | 1.0 | -2.475 |
| 33 | 5 | p | -0.5 | -2.0 | -2.475 |
| 34 | 5 | d | -0.5 | -1.0 | -2.475 |
| 35 | 4 | d | -0.5 | 0.0 | -2.475 |
| 36 | 5 | d | -0.5 | 1.0 | -2.475 |
| 37 | 5 | p | -0.5 | 2.0 | -2.475 |
| 38 | 5 | p | 0.5 | -2.0 | -2.475 |
| 39 | 5 | d | 0.5 | -1.0 | -2.475 |
| 40 | 4 | d | 0.5 | 0.0 | -2.475 |
| 41 | 5 | d | 0.5 | 1.0 | -2.475 |
| 42 | 5 | p | 0.5 | 2.0 | -2.475 |
| 43 | 6 | f | 1.5 | -1.0 | -2.475 |
| 44 | 4 | p | 1.5 | 0.0 | -2.475 |
| 45 | 6 | f | 1.5 | 1.0 | -2.475 |
| 46 | 6 | d | 2.5 | -1.0 | -2.475 |
| 47 | 4 | s | 2.5 | 0.0 | -2.475 |
| 48 | 6 | d | 2.5 | 1.0 | -2.475 |
| 49 | 6 | f | -2.0 | -1.5 | -1.768 |
| 50 | 4 | p | -2.0 | -0.5 | -1.768 |

**Tab 1.**
page 1

| Nucleon number | Corresponding Period | Sub-shell | Coordinates X | Y | Z |
|---|---|---|---|---|---|
| 51 | 4 | p | -2.0 | 0.5 | -1.768 |
| 52 | 6 | f | -2.0 | 1.5 | -1.768 |
| 53 | 5 | d | -1.0 | -1.5 | -1.768 |
| 54 | 4 | d | -1.0 | -0.5 | -1.768 |
| 55 | 4 | d | -1.0 | 0.5 | -1.768 |
| 56 | 5 | d | -1.0 | 1.5 | -1.768 |
| 57 | 3 | s | 0.0 | -1.5 | -1.768 |
| 58 | 3 | p | 0.0 | -0.5 | -1.768 |
| 59 | 3 | p | 0.0 | 0.5 | -1.768 |
| 60 | 3 | s | 0.0 | 1.5 | -1.768 |
| 61 | 5 | d | 1.0 | -1.5 | -1.768 |
| 62 | 4 | d | 1.0 | -0.5 | -1.768 |
| 63 | 4 | d | 1.0 | 0.5 | -1.768 |
| 64 | 5 | d | 1.0 | 1.5 | -1.768 |
| 65 | 6 | f | 2.0 | -1.5 | -1.768 |
| 66 | 4 | p | 2.0 | -0.5 | -1.768 |
| 67 | 4 | p | 2.0 | 0.5 | -1.768 |
| 68 | 6 | f | 2.0 | 1.5 | -1.768 |
| 69 | 4 | d | -1.5 | -1.0 | -1.061 |
| 70 | 2 | s | -1.5 | 0.0 | -1.061 |
| 71 | 4 | d | -1.5 | 1.0 | -1.061 |
| 72 | 3 | p | -0.5 | -1.0 | -1.061 |
| 73 | 2 | p | -0.5 | 0.0 | -1.061 |
| 74 | 3 | p | -0.5 | 1.0 | -1.061 |
| 75 | 3 | p | 0.5 | -1.0 | -1.061 |
| 76 | 2 | p | 0.5 | 0.0 | -1.061 |
| 77 | 3 | p | 0.5 | 1.0 | -1.061 |
| 78 | 4 | d | 1.5 | -1.0 | -1.061 |
| 79 | 2 | s | 1.5 | 0.0 | -1.061 |
| 80 | 4 | d | 1.5 | 1.0 | -1.061 |
| 81 | 2 | p | -1.0 | -0.5 | -0.354 |
| 82 | 2 | p | -1.0 | 0.5 | -0.354 |
| 83 | 1 | s | 0.0 | -0.5 | -0.354 |
| 84 | 1 | s | 0.0 | 0.5 | -0.354 |
| 85 | 2 | p | 1.0 | -0.5 | -0.354 |
| 86 | 2 | p | 1.0 | 0.5 | -0.354 |
| 87 | 2 | p | -0.5 | -1.0 | 0.354 |
| 88 | 1 | s | -0.5 | 0.0 | 0.354 |
| 89 | 2 | p | -0.5 | 1.0 | 0.354 |
| 90 | 2 | p | 0.5 | -1.0 | 0.354 |
| 91 | 1 | s | 0.5 | 0.0 | 0.354 |
| 92 | 2 | p | 0.5 | 1.0 | 0.354 |
| 93 | 4 | d | -1.0 | -1.5 | 1.061 |
| 94 | 3 | p | -1.0 | -0.5 | 1.061 |
| 95 | 3 | p | -1.0 | 0.5 | 1.061 |
| 96 | 4 | d | -1.0 | 1.5 | 1.061 |
| 97 | 2 | s | 0.0 | -1.5 | 1.061 |
| 98 | 2 | p | 0.0 | -0.5 | 1.061 |
| 99 | 2 | p | 0.0 | 0.5 | 1.061 |
| 100 | 2 | s | 0.0 | 1.5 | 1.061 |

**Tab 1.**
page 2

| Nucleon | Corresponding | | | Coordinates | |
|---|---|---|---|---|---|
| number | Period | Sub-shell | X | Y | Z |
| 101 | 4 | d | 1.0 | -1.5 | 1.061 |
| 102 | 3 | p | 1.0 | -0.5 | 1.061 |
| 103 | 3 | p | 1.0 | 0.5 | 1.061 |
| 104 | 4 | d | 1.0 | 1.5 | 1.061 |
| 105 | 6 | f | -1.5 | -2.0 | 1.768 |
| 106 | 5 | d | -1.5 | -1.0 | 1.768 |
| 107 | 3 | s | -1.5 | 0.0 | 1.768 |
| 108 | 5 | d | -1.5 | 1.0 | 1.768 |
| 109 | 6 | f | -1.5 | 2.0 | 1.768 |
| 110 | 4 | p | -0.5 | -2.0 | 1.768 |
| 111 | 4 | d | -0.5 | -1.0 | 1.768 |
| 112 | 3 | p | -0.5 | 0.0 | 1.768 |
| 113 | 4 | d | -0.5 | 1.0 | 1.768 |
| 114 | 4 | p | -0.5 | 2.0 | 1.768 |
| 115 | 4 | p | 0.5 | -2.0 | 1.768 |
| 116 | 4 | d | 0.5 | -1.0 | 1.768 |
| 117 | 3 | p | 0.5 | 0.0 | 1.768 |
| 118 | 4 | d | 0.5 | 1.0 | 1.768 |
| 119 | 4 | p | 0.5 | 2.0 | 1.768 |
| 120 | 6 | f | 1.5 | -2.0 | 1.768 |
| 121 | 5 | d | 1.5 | -1.0 | 1.768 |
| 122 | 3 | s | 1.5 | 0.0 | 1.768 |
| 123 | 5 | d | 1.5 | 1.0 | 1.768 |
| 124 | 6 | f | 1.5 | 2.0 | 1.768 |
| 125 | 5 | p | -2.0 | -0.5 | 2.475 |
| 126 | 5 | p | -2.0 | 0.5 | 2.475 |
| 127 | 6 | d | -1.0 | -2.5 | 2.475 |
| 128 | 6 | f | -1.0 | -1.5 | 2.475 |
| 129 | 5 | d | -1.0 | -0.5 | 2.475 |
| 130 | 5 | d | -1.0 | 0.5 | 2.475 |
| 131 | 6 | f | -1.0 | 1.5 | 2.475 |
| 132 | 6 | d | -1.0 | 2.5 | 2.475 |
| 133 | 4 | s | 0.0 | -2.5 | 2.475 |
| 134 | 4 | p | 0.0 | -1.5 | 2.475 |
| 135 | 4 | d | 0.0 | -0.5 | 2.475 |
| 136 | 4 | d | 0.0 | 0.5 | 2.475 |
| 137 | 4 | p | 0.0 | 1.5 | 2.475 |
| 138 | 4 | s | 0.0 | 2.5 | 2.475 |
| 139 | 6 | d | 1.0 | -2.5 | 2.475 |
| 140 | 6 | f | 1.0 | -1.5 | 2.475 |
| 141 | 5 | d | 1.0 | -0.5 | 2.475 |
| 142 | 5 | d | 1.0 | 0.5 | 2.475 |
| 143 | 6 | f | 1.0 | 1.5 | 2.475 |
| 144 | 6 | d | 1.0 | 2.5 | 2.475 |
| 145 | 5 | p | 2.0 | -0.5 | 2.475 |
| 146 | 5 | p | 2.0 | 0.5 | 2.475 |
| 147 | 5 | s | -2.5 | 0.0 | 3.182 |
| 148 | 5 | p | -1.5 | 0.0 | 3.182 |
| 149 | 6 | p | -0.5 | -3.0 | 3.182 |
| 150 | 6 | d | -0.5 | -2.0 | 3.182 |

**Tab 1.**
page 3

| Nucleon | Corresponding | | | Coordinates | |
|---|---|---|---|---|---|
| number | Period | Sub-shell | X | Y | Z |
| 151 | 6 | f | -0.5 | -1.0 | 3.182 |
| 152 | 5 | d | -0.5 | 0.0 | 3.182 |
| 153 | 6 | f | -0.5 | 1.0 | 3.182 |
| 154 | 6 | d | -0.5 | 2.0 | 3.182 |
| 155 | 6 | p | -0.5 | 3.0 | 3.182 |
| 156 | 6 | p | 0.5 | -3.0 | 3.182 |
| 157 | 6 | d | 0.5 | -2.0 | 3.182 |
| 158 | 6 | f | 0.5 | -1.0 | 3.182 |
| 159 | 5 | d | 0.5 | 0.0 | 3.182 |
| 160 | 6 | f | 0.5 | 1.0 | 3.182 |
| 161 | 6 | d | 0.5 | 2.0 | 3.182 |
| 162 | 6 | p | 0.5 | 3.0 | 3.182 |
| 163 | 5 | p | 1.5 | 0.0 | 3.182 |
| 164 | 5 | s | 2.5 | 0.0 | 3.182 |
| 165 | 6 | s | 0.0 | -3.5 | 3.889 |
| 166 | 6 | p | 0.0 | -2.5 | 3.889 |
| 167 | 6 | d | 0.0 | -1.5 | 3.889 |
| 168 | 6 | f | 0.0 | -0.5 | 3.889 |
| 169 | 6 | f | 0.0 | 0.5 | 3.889 |
| 170 | 6 | d | 0.0 | 1.5 | 3.889 |
| 171 | 6 | p | 0.0 | 2.5 | 3.889 |
| 172 | 6 | s | 0.0 | 3.5 | 3.889 |

**Tab. S1** Ideal coordinates of the nucleons of noble gases. The radius of the nucleons is 0.5 unit. The origin of the coordinate system is the same as the center of the mass of the double tetrahedron. The periods correspond to the periods of the periodic table.

**Tab 1.**
page 4

**Tab. S 1**